\documentclass[a4paper,final]{rqaa}

\usepackage{amsopn}
\usepackage[final]{graphicx}
\usepackage{natbib}
\usepackage{txfonts}

\newcommand{\mat}{\vec}
\newcommand{\transpose}{\mathrm{T}}

\begin{document}

\title{Evolution strategies applied to the problem of line profile
  decomposition in QSO spectra}
\author{R.~Quast \and
  R.~Baade \and
  D.~Reimers}
\authorrunning{R.~Quast et al.}
\institute{Hamburger Sternwarte, Universit\"at Hamburg,
  Gojenbergsweg 112, D-21029 Hamburg, Germany\\
  \email{\{rbaade,dreimers\}@hs.uni-hamburg.de}}
\offprints{R.~Quast\\
  \email{rquast@hs.uni-hamburg.de}}
\date{Received July 6, 2004 / Accepted September 16, 2004}
\idline{To appear in Astron. Astrophys.}{1}

\defcitealias{HansenN_OstermeierA_2001}{Paper~A}
\defcitealias{HansenMK_2003}{Paper~B}
\defcitealias{PressTVF_2002}{Numerical Recipes}

\abstract{%
We describe the decomposition of QSO absorption line ensembles applying an
evolutionary forward modelling technique. The modelling is optimized 
using an evolution strategy (ES) based on a novel concept of completely
derandomized self-adaption. The algorithm is described in detail. Its
global optimization performance in decomposing a series of simulated test
spectra is compared to that of classical deterministic algorithms. Our
comparison demonstrates that the ES is a highly competitive algorithm
capable to calculate the optimal decomposition without requiring any
particular initialization.

}

\maketitle

\section{Introduction} 

The standard astronomical data analysis packages such as the Image
Reduction and Analysis Facility (IRAF) and many popular custom-built
applications offer only deterministic algorithms for the purpose of
parametric model fitting. In practice, however, deterministic algorithms
often require considerable operational intervention by the user. In
contrast, stochastic strategies such as evolutionary algorithms minimize
the operational interaction, a highly appealing feature.

In practice, any parametric model fitting technique reduces to the
numerical problem of finding the minimum of an objective function
$f:\mathbb{R}^n\to\mathbb{R}$, i.e. constructing a sequence of object
parameter vectors
\begin{equation}
  (\vec{x}^{(g)})_{g\in\mathbb{N}},
  \quad
  \vec{x}^{(g)}\in\mathbb{R}^n
  \label{eq:sequence}
\end{equation}
such that $\lim_{g\to\infty}f(\vec{x}^{(g)})$ is as small as
possible. Several classical algorithms are practicable to overcome the
problem of minimization. Among the best established strategies are the
conjugate gradient, variable metric, and Levenberg-Marquardt algorithms,
all gathering information about the local topography of $f$ by calculating
its partial derivatives (i.e. the gradient or the Hessian matrix) and
thereby ensuring the rapid convergence at a close minimum
\citep[e.g.,][Numerical Recipes]{PressTVF_2002}. The
nature of these classical algorithms is deterministic: each member of the
sequence $(\vec{x}^{(g)})_g$ is determined by its predecessor, i.e.
the whole sequence is determined by the intial object parameter vector
$\vec{x}^{(0)}$. Exactly this fact turns out to be the cardinal deficiency
in case $f$ exhibits many local minima: The success of the algorithms in
locating the global minimum of $f$ crucially depends on the adequacy of
the initial guess $\vec{x}^{(0)}$. Hence, the results are biased and
several optimization runs are normally necessary. In particular in case of
noisy data or strong inter-parameter correlations the problem of finding
an adequate initialization $\vec{x}^{(0)}$ is tedious and often dominates
the time needed to complete the optimization. In addition, the classical
algorithms are not practicable if the objective function is not
continuously differentiable, oscillating, or the calculation of partial
derivatives is too expensive or numerically inaccurate.

In contrast, stochastic strategies such as evolutionary algorithms where
each member of the sequence $(\vec{x}^{(g)})_g$ is the result of a
random experiment, are not only promising for the purpose of global
optimization but also apply to any objective function which is computable
or available by experiment. However, the drawback
of stochastic optimization strategies is less efficiency: even if an
adequate local downhill move exists, a random step will almost always
lead astray. Therefore, many more evaluations of the objective function
are required before the sequence $(\vec{x}^{(g)})_g$ converges at a
minimum.

Evolutionary algorithms are inspired by the principles of biological
evolution. Conceptually, three major subclasses are discerned:
evolutionary programming, genetic algorithms, and evolution strategies
(ES). While all subclasses apply quite generally, ES are particularly
suited for the purpose of continuous parametric  optimization. Largely
\citet[hereafter Paper~A]{HansenN_OstermeierA_2001} and
\citet[hereafter Paper~B]{HansenMK_2003} have established a novel concept
of completely derandomized self-adaption in ES which selectively
approximates the inverse Hessian matrix of the objective function and
thereby considerably improves the efficiency in case of non-separable
problems or mis-scaled parameter mappings while even increasing the chance
of finding the global optimum in case of strong multimodality.

The interpretation and analysis of QSO absorption lines basically involves
the decomposition of line ensembles into individual line profiles. In
general, the decomposition of QSO spectra presents an ambigous parametric
inverse problem and automatizing the decomposition demands an efficient
but at first stable optimization algorithm. In this study we resume the
concept of completely derandomized self-adaption introduced in
\citetalias{HansenN_OstermeierA_2001} and test the global
optimization performance of the resulting ES when applied to the problem
of line profile decomposition in QSO spectra.

\section{Evolution strategies (ES)} 

\subsection{General concepts}

The state of an ES in generation $g$ is defined by the parental family of
object parameter vectors $\vec{x}^{(g)}_1,\dots,
\vec{x}^{(g)}_{\mu\in\mathbb{N}}\in\mathbb{R}^n$ and the mutation
operator $p^{(g)}:\mathbb{R}^n\to\mathbb{R}^n$. The generic ES algorithm
is completely defined by the transition from generation $g$ to $g+1$. For
instance, in the illustrative case of a simple single parent strategy:
\begin{enumerate}
\item 
Mutation of the parental vector $\vec{x}^{(g)}$ to produce a new
population of $\lambda>1$ offsprings
\begin{equation}
  \vec{y}^{(g+1)}_1,\dots,\vec{y}^{(g+1)}_\lambda\in\mathbb{R}^n
    \quad
    \leftarrow
    \quad
     p^{(g)}\vec{x}^{(g)} = N(\vec{x}^{(g)},\sigma^2\mat{I}),
\end{equation}
where each offspring is sampled from an $n$-dimensional normal mutation
distribution with mean $\vec{x}^{(g)}$ and isotropic variance $\sigma^2$.
\item 
Selection of the best individual among the offspring population to become
the new parental vector
\begin{equation}
  \vec{x}^{(g+1)} =
    \vec{y}_{1:\lambda}^{(g+1)},
\end{equation}
where the notation $\vec{y}_{i:\lambda}$ refers to the $i^\mathrm{th}$
best among $\vec{y}_1,\dots,\vec{y}_\lambda$ individuals, and $\vec{y}_i$
is better than $\vec{y}_j$ if $f(\vec{y}_i)<f(\vec{y}_j)$.
\end{enumerate}

In general, the transition of the parental family of object parameter
vectors $\vec{x}^{(g)}_1,\dots,\vec{x}^{(g)}_\mu$ from generation $g$ to
$g+1$ is accomplished according to the following coherent scheme:
\begin{enumerate}
\item 
Recombination of $\rho\le\mu$ randomly selected parental vectors into the
recombinant vector $\langle\vec{x}\rangle^{(g)}$. The recombination is
either a definite or random algebraic operation.
\item 
Mutation of the recombinant vector $\langle\vec{x}\rangle^{(g)}$ to
produce a new population of $\lambda>\mu$ offsprings
\begin{equation}
  \vec{y}^{(g+1)}_1,\dots,\vec{y}^{(g+1)}_\lambda\in\mathbb{R}^n
    \quad
    \leftarrow
    \quad
    p^{(g)}\langle\vec{x}\rangle^{(g)},
\end{equation}
where the maximally unbiased mutation operator corresponds to an
$n$-dimensional correlated normal mutation distribution with zero mean and
covariance matrix $\mat{C}^{(g)}$, i.e
\begin{equation}
  p^{(g)}\langle\vec{x}\rangle^{(g)} =
    \langle\vec{x}\rangle^{(g)} + N(0,\mat{C}^{(g)}).
\end{equation}
\item 
Selection of the $\mu$ best individuals among the offspring population (or
the union of offsprings and parents) to become the next generation of
parental vectors
\begin{equation}
  \vec{x}^{(g+1)}_i =
    \vec{y}^{(g+1)}_{i:\lambda},
    \quad
    i=1,\dots,\mu.
\end{equation}
\end{enumerate}

\begin{figure}
  \includegraphics[width=\columnwidth]{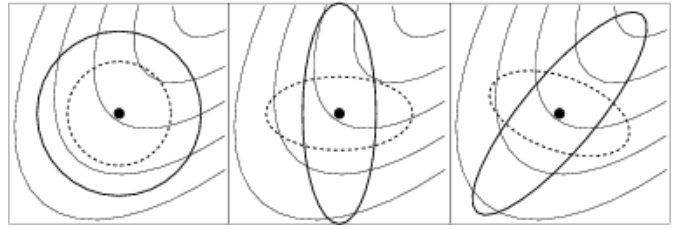}
  \caption[]{Examples of isotropic, uncorrelated, and correlated
    mutation distributions (indicated by circles,
    axis-parallel ellipsoids, and rotated ellipsoids, respectively)
    rendered over an elongated valley topography. The best average
    progress toward the topographic minimum in direction of the upper
    right corner is achieved in the right panel, where the mutation
    distribution (solid ellipsoid) is adapted to the topography}
  \label{fg:ellipsoids}
\end{figure}

In contrast to the transition of parental vectors, there is no coherent
conceptual scheme for the transition of the mutation operator $p^{(g)}$
from generation $g$ to $g+1$. However, due to basic considerations,
any advanced transition scheme is expected to reflect the following
elementary principles:
\begin{enumerate}
\item 
Invariance of the resulting ES with respect to any strictly monotone
remapping of the range of the objective function as well as any linear
transformation of the object parameter space. 
In particular, the resulting ES is expected to be unaffected by
translation, rotation, and reflection. 
\item 
Self-adaption of (the shape of) the mutation distribution to the
topography of the objective function (Fig.~\ref{fg:ellipsoids}). In
particular, the mutation distribution is expected to reproduce the
precedently selected mutation steps with increased likelihood.
\end{enumerate}

The rigid implementation of these principles results in a concept of
completely derandomized self-adaption where the transition of the
mutation operator from one generation to the next is accomplished by
successively updating the covariance matrix of the mutation distribution
with information provided by the actually selected mutation step. The
further demand of nonlocality involves the cumulation of the selected
mutation steps into an evolution path.

\subsection{Covariance matrix adaption (CMA)}

\begin{figure}
  \includegraphics[width=\columnwidth]{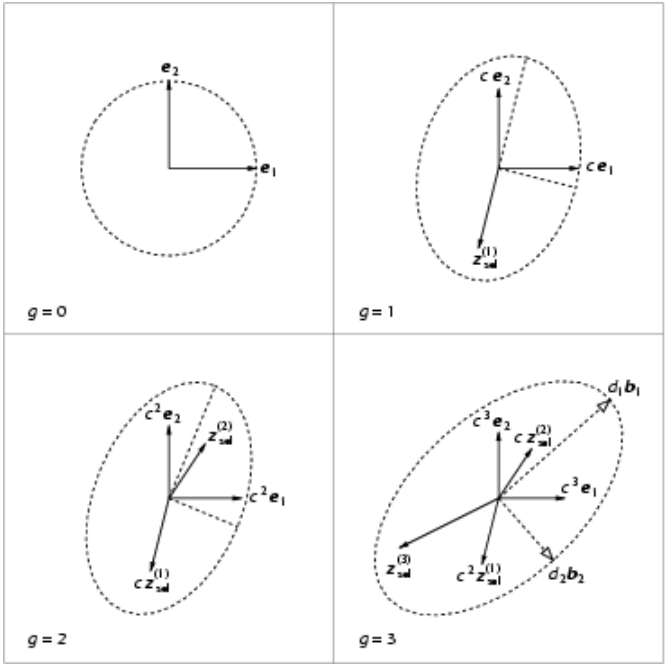}
  \caption[]{The conceptual scheme of covariance matrix adaption (CMA).
    Initially, the covariance matrix of the mutation distribution is
    $\mat{C}^{(0)}=\vec{e}_1\vec{e}_1^\transpose+
    \vec{e}_2\vec{e}_2^\transpose$.
    In the course of the transition to the next generation $g+1$ the
    symmetric rank-one matrix $\vec{z}^{(g+1)}_\mathrm{sel}
    (\vec{z}^{(g+1)}_\mathrm{sel})^\transpose$,
    where $\vec{z}^{(g+1)}_\mathrm{sel}$ is the actually selected mutation
    step, is added to the downscaled covariance matrix. In the third
    generation, the covariance matrix is
    $\mat{C}^{(3)}=c^6\vec{e}_1
    \vec{e}_1^\transpose+c^6\vec{e}_2\vec{e}_2^\transpose+
    \sum_{i=1}^3c^{2(3-i)}\vec{z}_\mathrm{sel}^{(i)}
    (\vec{z}_\mathrm{sel}^{(i)})^\transpose=d_1^2\vec{b}_1
    \vec{b}_1^\transpose+d_2^2\vec{b}_2\vec{b}_2^\transpose$, where
    $\vec{b}_1$ and $\vec{b}_2$ are the unit eigenvectors of
    $\mat{C}^{(3)}$ corresponding to the eigenvalues $d_1^2$ and
    $d_2^2$. The resulting mutation distribution reads
    $N(0,\mat{C}^{(3)})=N(0,1)\,d_1\vec{b}_1+N(0,1)\,
    d_2\vec{b}_2$}
  \label{fg:cma}
\end{figure}

If $\vec{z}_1,\dots,\vec{z}_{m\in\mathbb{N}}\in\mathbb{R}^n$, $m\geq n$
are a generating set of $\mathbb{R}^n$ and $q_1,\dots,q_m\in\mathbb{R}$
are a sequence of $(0,1)$-normally distributed random variables, then
\begin{equation}
  q_1\vec{z}_1 + \dots + q_m\vec{z}_m
  \label{eq:lc}
\end{equation}
renders an $n$-dimensional normal distribution with zero mean and
covariance matrix
\begin{equation}
  \vec{z}_1\vec{z}_1^\transpose + \dots +
    \vec{z}_m\vec{z}_m^\transpose.
\end{equation}
In fact, Eq.~(\ref{eq:lc}) facilitates the realization of any normal
distribution with zero mean. In particular, the symmetric rank-one
matrix $\vec{z}_i\vec{z}_i^\transpose$ corresponds to that normal
distribution with zero mean producing the vector $\vec{z}_i$ with the
maximum likelihood.
Since the objective of CMA is to reproduce the precedently selected
mutation steps with increased likelihood, the whole trick in order to
accomplish this objective is to add the symmetric rank-one matrix
\begin{equation}
  \vec{z}^{(g+1)}_\mathrm{sel}(\vec{z}^{(g+1)}_\mathrm{sel})^\transpose,
  \label{eq:r1m}
\end{equation}
where $\vec{z}^{(g+1)}_\mathrm{sel}$ denotes the actually selected
mutation step, to the covariance matrix of the mutation distribution. The
conceptual scheme is illustrated in Fig.~\ref{fg:cma}. Initially, the
mutation distribution is isotropic
\begin{equation}
  N(0,\mat{C}^{(0)}) =
    N(0,1)\,\vec{e}_1 + N(0,1)\,\vec{e}_2,
\end{equation}
whereas in the course of the transition to the next generation the
symmetric rank-one matrix Eq.~(\ref{eq:r1m}) is added to the downscaled
covariance matrix
\begin{equation}
  \mat{C}^{(g+1)} =
    c^2\mat{C}^{(g)} +
    \vec{z}_\mathrm{sel}^{(g+1)}(\vec{z}_\mathrm{sel}^{(g+1)})^\transpose,
    \quad
    c\in[0,1).
\end{equation}
In the course of the third generation, for instance, the mutation
distribution reads
\begin{eqnarray}
   N(0,\mat{C}^{(3)})
   & =
   &
     N(0,1)\,c^3\vec{e}_1 +
     N(0,1)\,c^3\vec{e}_2 +
     \sum_{i=1}^3 N(0,1)\,c^{3-i}\vec{z}_\mathrm{sel}^{(i)} \nonumber\\
   & =
   &
     N(0,1)\,d_1\vec{b}_1 +
     N(0,1)\,d_2\vec{b}_2,
  \label{eq:cma}
\end{eqnarray}
where $\vec{b}_1$ and $\vec{b}_2$ are the unit eigenvectors of
$\mat{C}^{(3)}$ corresponding to the eigenvalues $d_1^2$ and $d_2^2$.
Obviously, the mutation distribution tends to reproduce the precedently
selected mutations steps. Finally, the mutation distribution becomes
stationary (apart from scale) while its principal axes achieve conjugate
perpendicularity, and $\mat{C}^{(g)}$ effectively approaches the scaled
inverse Hessian matrix of the objective function.

\subsection{Evolution path cumulation}

The efficiency as well as the stability of an ES improve significantly,
if the decision of how to adapt the mutation distribution is based on the
cumulation of several selected mutation steps rather than a single step.
While the latter maximizes the local selection probability the former is
more likely to advance the global progress rate. The benefit from
superseding the successively selected mutation steps in favor of the
evolution path
\begin{equation}
  \vec{p}^{(g+1)} =
    (1-c)\vec{p}^{(g)} + \sqrt{c(2-c)} \vec{z}_\mathrm{sel}^{(g+1)},
  \quad
  c\in(0,1]
\end{equation}
is illustrated in detail in
\citetalias{HansenN_OstermeierA_2001}: If several successively selected
mutation steps are parallel (antiparallel) correlated, the evolution path
will be lengthened (shortened). If the evolution path is long (short), the
size of the mutation steps in direction of the evolution path increases
(decreases). The effect of cumulation is particularly beneficial in the
case of small populations where the topographical information gathered
within one generation is not sufficient. If $c=1$, no cumulation will
occur and $\vec{p}^{(g+1)}=\vec{z}_\mathrm{sel}^{(g+1)}$.

\subsection{The CMA evolution strategy (CMA-ES)}

In this section we compile a unified formulation of the generic CMA-ES
algorithms given in Papers A and B. In this form, the algorithm is not
presented elsewhere. Besides the rank-one CMA outlined in the previous
sections the algorithm features the advanced-rank CMA introduced in
\citetalias{HansenMK_2003} and an additional step size control. Finally,
we conclude with remarks concerning the numerical implementation of the
CMA-ES algorithm we provide online.

\subsubsection{Generic algorithm}

The state of the $(\mu,\lambda)$-CMA-ES in generation $g$ is defined by
the family of parental vectors $\vec{x}_1^{(g)},\dots,
\vec{x}_\mu^{(g)}\in\mathbb{R}^n$, the covariance matrix of the mutation
distribution $\mat{C}^{(g)}\in\mathbb{R}^{n\times n}$, the global mutation
step size $\sigma^{(g)}\in\mathbb{R}^+$, and the evolution paths
$\vec{p}_\mathrm{c}^{(g)},\vec{p}_\sigma^{(g)}\in\mathbb{R}^n$. The
generic algorithm is completely defined by the transition from generation
$g$ to $g+1$:
\begin{enumerate}
\item 
Weighted intermediate recombination\footnote{Discrete recombination such
as the cross-over commonly practiced in genetic algorithms is not
invariant with respect to linear transformations of the search space.}
of the parental vectors into the recombinant vector
\begin{equation}
  \langle\vec{x}\rangle^{(g)} =
    \frac{\sum_{i=1}^\mu w_i\vec{x}_i^{(g)}}{\sum_{i=1}^\mu w_i},
    \quad
    w_1,\dots,w_\mu\in\mathbb{R}^+.
  \label{eq:xw}
\end{equation}
The weights $w_1,\dots,w_\mu$ are internal strategy parameters
with canonical values
\begin{equation}
  w_i =
    \ln(\mu+1) - \ln(i),
    \quad
    i=1,\dots,\mu.
\end{equation}
The initial recombinant vector $\langle\vec{x}\rangle^{(0)}$ is expected
to enable the resulting mutation operator to sample the relevant part of
the object parameter space.
\item 
Mutation of the recombinant vector to produce a new family of
$\lambda>\mu$ offsprings
\begin{equation}
  \vec{y}_k^{(g+1)} =
    \langle\vec{x}\rangle^{(g)} +
    \sigma^{(g)} \mat{B}^{(g)}\mat{D}^{(g)}\vec{z}_k^{(g+1)},
    \quad
    k=1,\dots,\lambda,
  \label{eq:xk}
\end{equation}
where $\vec{z}_1^{(g+1)},\dots,\vec{z}_\lambda^{(g+1)}\in\mathbb{R}^n$
are a family of $(0,\mat{I})$-normally distributed random
vectors (i.e. the vector components are $(0,1)$-normally distributed),
and $\mat{D}^{(g)},\mat{B}^{(g)}\in\mathbb{R}^{n\times n}$ diagonalize
the covariance matrix
\begin{equation}
  \mat{C}^{(g)} =
    \mat{B}^{(g)}\mat{D}^{(g)}\mat{D}^{(g)}(\mat{B}^{(g)})^\transpose.
  \label{eq:evd}
\end{equation}
The elements of the diagonal matrix $\mat{D}^{(g)}$ are the square roots
of the eigenvalues of $\mat{C}^{(g)}$, while the column vectors of
$\mat{B}^{(g)}$ are the corresponding unit eigenvectors. Constrained
optimization can be realized by resampling $\vec{z}_1^{(g+1)},\dots,
\vec{z}_\lambda^{(g+1)}$ until the constraints are satisfied.
\item 
Selection of the $\mu$ best individuals among the offspring population to
become the next generation of parental vectors
\begin{equation}
  \vec{x}^{(g+1)}_i =
    \vec{y}^{(g+1)}_{i:\lambda},
    \quad
    i=1,\dots,\mu.
\end{equation}
Since the production of offsprings is independent of order the values of
the objective function
\begin{equation}
  f(\vec{y}_k^{(g+1)}),\quad k=1,\dots,\lambda
\end{equation}
can be computed in parallel.
\item 
Cumulation of the distribution evolution path
\begin{equation}
  \vec{p}_\mathrm{c}^{(g+1)} =
    (1-c_\mathrm{c})\vec{p}_\mathrm{c}^{(g)} +
      c_w\sqrt{c_\mathrm{c}(2-c_\mathrm{c})}
       \mat{B}^{(g)}\mat{D}^{(g)} \langle\vec{z}\rangle^{(g+1)}
  \label{eq:pc}
\end{equation}
with
\begin{equation}
  c_w =
    \frac{\sum_{i=1}^\mu w_i}{\sqrt{\sum_{i=1}^\mu w_i^2}}
  \label{eq:cw}
\end{equation}
and
\begin{equation}
  \langle\vec{z}\rangle^{(g+1)} =
    \frac{\sum_{i=1}^\mu
      w_i\vec{z}^{(g+1)}_{i:\lambda}}{\sum_{i=1}^\mu w_i},
  \label{eq:zw}
\end{equation}
where $\vec{z}^{(g+1)}_{i:\lambda}$ follows from
$\vec{y}^{(g)}_{i:\lambda}$ by Eq.~(\ref{eq:xk}). The distribution
cumulation rate $c_\mathrm{c}\in(0,1]$ is an internal strategy
parameter with canonical value
\begin{equation}
  c_\mathrm{c} =
    \frac{4}{n+4}.
\end{equation}
If $c_\mathrm{c}=1$, no cumulation will occur. Initially, the
distribution evolution path is $\vec{p}_\mathrm{c}^{(0)}=0$.
\item 
Adaption of the covariance matrix
\begin{eqnarray}
  \mat{C}^{(g+1)}
  &
  =
  & (1-c_\mathrm{cov})\,\mat{C}^{(g)} + c_\mathrm{cov}
    \left(
      \alpha_\mathrm{cov}\vec{p}_\mathrm{c}^{(g+1)}
      (\vec{p}_\mathrm{c}^{(g+1)})^\transpose
    \right.\nonumber\\
  &
  &
   \phantom{(1-c_\mathrm{cov})\,\mat{C}^{(g)} + c_\mathrm{cov}[} +
     \left.
      (1-\alpha_\mathrm{cov})\langle\mat{Z}\rangle^{(g+1)}
    \right),
  \label{eq:cmatrix}
\end{eqnarray}
where
\begin{equation}
  \langle\mat{Z}\rangle^{(g+1)} =
    \frac{\sum_{i=1}^\mu
      w_i\mat{B}^{(g)}\mat{D}^{(g)}\vec{z}_{i:\lambda}^{(g+1)}
      \left(
        \mat{B}^{(g)}\mat{D}^{(g)}\vec{z}_{i:\lambda}^{(g+1)}
      \right)^\transpose}{\sum_{i=1}^\mu w_i}.
  \label{eq:zmatrix}
\end{equation}
The expressions
$\vec{p}_\mathrm{c}^{(g+1)}(\vec{p}_\mathrm{c}^{(g+1)})^\transpose$ and
$\langle\mat{Z}\rangle^{(g+1)}$ are symmetric matrices of rank one and
$\min(\mu,n)$, respectively, and generalize the conceptual scheme
illustrated in Fig.~\ref{fg:cma} in case of a single parent ES. The
parameter $\alpha_\mathrm{cov}\in[0,1]$ combines the two adaption
mechanisms whereas $c_\mathrm{cov}\in[0,1)$ moderates the adaption rate.
Both numbers are internal strategy parameters with canonical values
\begin{equation}
  \alpha_\mathrm{cov} =
    c_w^{-2}
\end{equation}
and
\begin{equation}
  c_\mathrm{cov} =
    \frac{2\alpha_\mathrm{cov}}{(n+\sqrt{2})^2} +
      (1-\alpha_\mathrm{cov})\,
      \operatorname{min}
        \left(
          1,
          \frac{2c_w^2-1}{(n+2)^2+c_w^2}
        \right).
\end{equation}
If $c_\mathrm{cov}=0$, no adaption will occur. Initially,
$\mat{C}^{(0)}=\mat{I}$ or the square of any diagonal matrix properly
scaling the optimization problem.
\item 
Cumulation of the step size evolution path
\begin{equation}
  \vec{p}_\sigma^{(g+1)} =
    (1-c_\sigma)\vec{p}_\sigma^{(g)} +
      c_w\sqrt{c_\sigma(2-c_\sigma)}\mat{B}^{(g)}
      \langle\vec{z}\rangle^{(g+1)},
  \label{eq:ps}
\end{equation}
where the step size cumulation rate $c_\sigma\in(0,1]$ is an internal
strategy parameter with canonical value
\begin{equation}
  c_\sigma =
    \frac{c_w^2+2}{n+c_w^2+3}.
\end{equation}
If $c_\sigma=1$, no cumulation will occur. Initially, the step size
evolution path is $\vec{p}_\sigma^{(0)}=0$.
\item 
Adaption of the global mutation step size
\begin{equation}
  \sigma^{(g+1)} =
    \sigma^{(g)}\,
      \exp
        \left(
          \frac{c_\sigma}{d_\sigma}
          \frac{\|\vec{p}_\sigma^{(g+1)}\|-E_n}{E_n}
        \right),
  \label{eq:stepsize}
\end{equation}
where the second fraction is the relative deviation of the length of
$\vec{p}_\sigma^{(g+1)}$ from its expected value in the case of no
selection pressure,
$E_n=\sqrt{2}\Gamma(\frac{n+1}{2})/\Gamma(\frac{n}{2})$, and the damping
constant $d_\sigma\geq c_\sigma$ is an internal strategy parameter with
canonical value
\begin{equation}
  d_\sigma =
    1 + c_\sigma + 
      2\operatorname{max}
        \left(
          0,
          \sqrt{\frac{c_w^2-1}{n+1}} - 1
        \right).
\end{equation}
The initial mutation step size $\sigma^{(0)}$ is expected to enable the
resulting mutation distribution to sample the relevant part of the object
parameter space.
\end{enumerate}

The impact and canonical setting of internal strategy parameters are
discussed in detail in \citetalias{HansenN_OstermeierA_2001}
and \citet{HansenN_KernS_2004}.

\subsubsection{Numerical implementation}

The numerical implementation of the CMA-ES algorithm is basically
straightforward.\footnote{Sophisticated examples are given on the home
page of Nikolaus Hansen at \texttt{http://www.bionik.tu-berlin.de}} We
coded different function templates for both unconstrained and constrained
optimization. The template instantiation requires a random number
generator and an eigenvalue decomposition algorithm to be supplied as
generic arguments. The numerical code is designed to perform in parallel
when running on shared memory multiprocessing architectures and has been
applied in practice by \citet{QuastBR_2002,QuastRL_2004} and
\citet{ReimersBQL_2003}.\footnote{The source code is publicly available on
the home page of RQ at \texttt{http://www.hs.uni-hamburg.de}}

Since an adequate random number generator producing high-dimensional
equidistribution is absolutely essential to ensure the ES practically
performs as expected in theory, we apply the Mersenne Twister
algorithm furnishing
equidistributed uniform deviates in up to 623 dimensions with period
$2^{19937}-1$ \citep{MatsumotoM_NishimuraT_1998}. The uniform deviates are
converted into normal deviates by means of the polar method.

The Linear Algebra Package provides several algorithms suitable for
diagonalizing the covariance matrix. For example, the routines
decomposing tridiagonal matrices into relatively robust representations
accurately complete the diagonalization of an $n\times n$ symmetric
tridiagonal matrix in $O(n^2)$ rather than the regular $O(n^3)$ arithmetic
operations \citep{DhillonI_ParlettB_2004}.
In particular, it is feasible to diagonalize the covariance matrix just
every $n/10$ generation to minimize the numerical overhead
\citepalias{HansenN_OstermeierA_2001}. For, say, $n>10\,000$ using
completely correlated mutation distributions will be impracticable and an
ES algorithm calculating just the tridiagonal covariance matrix elements
or adapting just individual step sizes ($\mat{B}^{(g)}=\mat{I}$,
$\mat{C}^{(g)}=\mat{D}^{(g)}\mat{D}^{(g)}$ diagonal) will be appropriate.

\section{Spectral decomposition}

In the case of pure line absorption, the observed spectral flux
$F(\lambda)$ is modelled as the product of the continuum background
and the instrumentally convoluted absorption term
\begin{equation}
  F(\lambda) =
    C(\lambda)\int P(\xi)
      \,\mathrm{e}^{-\tau(\lambda-\xi)}\,\mathrm{d}\xi.
  \label{eq:flux1}
\end{equation}
Defining
\begin{equation}
  A(\lambda) =
    \int P(\xi)\,\mathrm{e}^{-\tau(\lambda-\xi)}\,\mathrm{d}\xi
  \label{eq:convolution}
\end{equation}
and approximating the local continuum background by a linear combination
of Legendre polynomials, Eq.~(\ref{eq:flux1}) transforms into
\begin{equation}
  F(\lambda) =
    \sum_k c_k\,L_k[\phi(\lambda)]A(\lambda),
  \label{eq:flux2}
\end{equation}
where $L_k$ denotes the Legendre polynomial of order $k$, and $\phi$ is a
linear map onto the interval $[-1,1]$. The instrumental profile $P(\xi)$
is modelled by a normalized Gaussian defined by the spectral resolution of
the instrument. The instrumental convolution can be calculated piecewise
analytically while approximating the absorption term with a polyline or a
cubic spline. Without significant loss of accuracy, the integration can be
restricted to the interval $|\xi|\le2\delta$, where $\delta$ is the
full width at half maximum of the instrumental profile.

Presuming pure Doppler broadening, the optical depth $\tau$ is
modelled by a superposition of Gaussian functions. If $\lambda_i$, $f_i$,
$z_i$, $b_i$, $N_i$, and $\lambda_{z_i}=(1+z_i)\lambda_i$ denote,
respectively, the rest wavelength, the oscillator strength, the
cosmological redshift, the line broadening velocity, the column density,
and the observed wavelength corresponding to line $i$, then
\begin{equation}
  \tau(\lambda) =
    \sum_i g_i(\lambda)
  \label{eq:tau}
\end{equation}
with
\begin{equation}
  g_i(\lambda) =
    \frac{e^2}{4\varepsilon_0 mc}
    \frac{N_i f_i \lambda_i}{\sqrt{\pi}b_i}\,
      \exp{
        \left[-
          \left(\frac{c}{b_i}
            \frac{\lambda - \lambda_{z_i}}{\lambda_{z_i}}
          \right)^2
        \right]}.
  \label{eq:gaussian}
\end{equation}
If natural broadening is important, the Gaussian functions will have to be
replaced with Voigt functions. The latter can efficiently be calculated by
using pseudo-Voigt approximations \citep{IdaAT_2000}.

Taking the proper identification of lines for granted, the decomposition
of an absorption line spectrum into individual profiles is a parametric
inverse problem involving a tuple of three model parameters per
line: position, broadening velocity, and column density. Given an
observed set of spectral fluxes $F_1,F_2,\dots$ and normally distributed
errors $\sigma_1,\sigma_2,\dots$ sampled at wavelengths $\lambda_1,
\lambda_2,\dots$ the optimal parametric decomposition $F(\lambda)$ is
calculated by minimizing the normalized residual sum of squares
\begin{equation}
  \mathrm{RSS} =
    \sum_j\left(
      \frac{F_j - F(\lambda_j)}{\sigma_j}
    \right)^2.
  \label{eq:rss1}
\end{equation}
For any $A(\lambda)$, the minimization with respect to the coefficients
$c_k$ presents a linear optimization problem
\begin{equation}
  \mathrm{RSS} =
    \sum_j\left(
      \frac{F_j -
        \sum_k c_k\,L_k[\phi(\lambda_j)]A(\lambda_j)}{\sigma_j}
    \right)^2
  \label{eq:rss2}
\end{equation}
which is solved directly by means of Cholesky decomposition. In the case
of normally distributed errors the minimum RSS maximizes  the likelihood.

Note that since the normalized RSS is invariant with respect to
permutations of line parameter 3-tuples, any self-adaptive ES initially
will require several generations to adapt to the symmetry. Any further
global degeneracies of the search space will lengthen the initial adaption
phase.

\section{Performance tests} 

\subsection{Test cases}

\begin{figure*}
  \centering
  \includegraphics{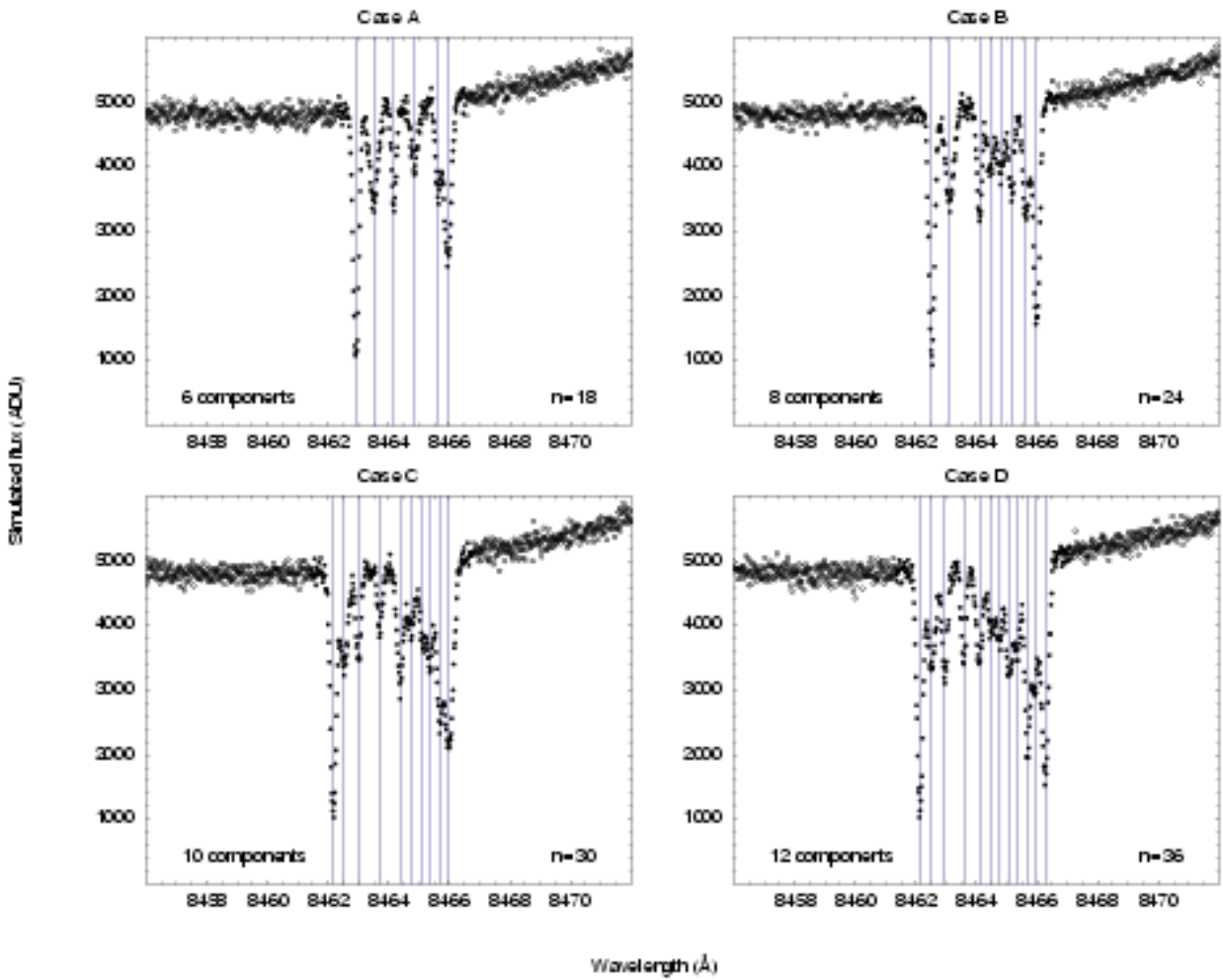}%
  \caption[]{Artificial test cases based on an ensemble of intergalactic
    \ion{Ca}{ii}~$\lambda3935$ lines toward the QSO
    HE~0515--4414. The simulated instrumental resolution is $R=60\,000$
    while the simulated signal-to-noise ratio is about 50. Individual
    components are marked by vertical lines. The positional search space
    is indicated by solid dots}
  \label{fg:cases}
\end{figure*}

\begin{table}
  \centering
  \caption[]{Parameterization of the artificial lines used to synthesize
    the test cases rendered in Fig.~\ref{fg:cases}. Rest wavelengths and
    oscillator strengths comply with the \ion{Ca}{ii}~$\lambda3935$
    transition.
    All CMA-ES runs converging at the global minimum result in the same
    parameterization indicated in parantheses, along with standard
    deviations provided by the diagonal elements of the scaled covariance
    matrix}
  \label{tb:cases}
  \begin{tabular}{@{}llll@{}}
    \hline
    \hline
    Line &
    $z$ &
    $b$ (km\,s$^{-1}$) &
    $\log N$ (cm$^{-2}$)\rule[-5pt]{0pt}{15pt}\\
    \hline
    A-1 & 
    1.150800 (800) & 
    3.0 ($2.9\pm0.1$) & 
    12.30 ($12.32\pm0.01$)\rule{0pt}{10pt}\\
    A-2 & 
    1.150950 (949) & 
    6.0 ($6.0\pm0.2$) & 
    11.80 ($11.80\pm0.01$)\\
    A-3 & 
    1.151120 (119) & 
    2.5 ($1.9\pm0.2$) & 
    11.60 ($11.60\pm0.02$)\\
    A-4 & 
    1.151290 (290) & 
    5.0 ($4.8\pm0.3$) & 
    11.50 ($11.51\pm0.02$)\\
    A-5 & 
    1.151490 (488) & 
    3.0 ($2.8\pm0.3$) & 
    11.60 ($11.58\pm0.02$)\\
    A-6 & 
    1.151570 (569) & 
    4.5 ($4.6\pm0.2$) & 
    12.00 ($12.01\pm0.01$)\rule[-5pt]{0pt}{10pt}\\
    \hline
    B-1 & 
    1.150700 (700) & 
    3.0 ($2.9\pm0.1$) & 
    12.30 ($12.32\pm0.01$)\rule{0pt}{10pt}\\
    B-2 & 
    1.150850 (850) & 
    6.0 ($6.0\pm0.2$) & 
    11.80 ($11.80\pm0.01$)\\
    B-3 & 
    1.151100 (101) & 
    2.5 ($2.5\pm0.2$) & 
    11.70 ($11.69\pm0.01$)\\
    B-4 & 
    1.151190 (189) & 
    4.0 ($4.0\pm0.5$) & 
    11.50 ($11.50\pm0.03$)\\
    B-5 & 
    1.151280 (279) &
     5.5 ($5.6\pm0.7$) & 
     11.60 ($11.61\pm0.03$)\\
    B-6 & 
    1.151370 (370) & 
    3.0 ($3.2\pm0.3$) & 
    11.60 ($11.60\pm0.02$)\\
    B-7 & 
    1.151490 (490) & 
    4.5 ($4.4\pm0.2$) & 
    11.80 ($11.80\pm0.01$)\\
    B-8 & 
    1.151580 (580) & 
    3.5 ($3.5\pm0.1$) & 
    12.20 ($12.20\pm0.01$)\rule[-5pt]{0pt}{10pt}\\
    \hline
    C-1 & 
    1.150610 (610) & 
    3.0 ($2.9\pm0.1$) & 
    12.30 ($12.32\pm0.02$)\rule{0pt}{10pt}\\
    C-2 & 
    1.150700 (701) & 
    5.5 ($5.4\pm0.3$) & 
    11.80 ($11.79\pm0.02$)\\
    C-3 & 
    1.150820 (821) & 
    3.0 ($2.8\pm0.3$) & 
    11.60 ($11.59\pm0.02$)\\
    C-4 & 
    1.151000 (000) & 
    2.5 ($2.5\pm0.4$) & 
    11.40 ($11.38\pm0.02$)\\
    C-5 & 
    1.151170 (170) & 
    4.0 ($3.9\pm0.3$) & 
    11.80 ($11.79\pm0.02$)\\
    C-6 & 
    1.151260 (261) & 
    5.5 ($5.9\pm0.8$) & 
    11.60 ($11.61\pm0.04$)\\
    C-7 & 
    1.151360 (358) & 
    3.5 ($3.2\pm0.8$) & 
    11.50 ($11.51\pm0.07$)\\ 
    C-8 & 
    1.151420 (421) & 
    4.0 ($3.9\pm0.8$) & 
    11.70 ($11.71\pm0.05$)\\ 
    C-9 & 
    1.151500 (500) & 
    3.5 ($3.3\pm0.4$) & 
    11.90 ($11.90\pm0.03$)\\
    C-10 & 
    1.151580 (579) & 
    6.0 ($6.1\pm0.2$) & 
    12.20 ($12.21\pm0.01$)\rule[-5pt]{0pt}{10pt}\\
    \hline
    D-1 & 
    1.150600 (600) & 
    3.0 ($3.0\pm0.1$) & 
    12.30 ($12.30\pm0.01$)\rule{0pt}{10pt}\\  
    D-2 & 
    1.150700 (700) & 
    5.5 ($5.5\pm0.2$) & 
    11.80 ($11.81\pm0.01$)\\
    D-3 & 
    1.150810 (810) & 
    3.0 ($3.0\pm0.2$) & 
    11.70 ($11.70\pm0.01$)\\ 
    D-4 & 
    1.150970 (968) & 
    2.5 ($2.5\pm0.2$) & 
    11.60 ($11.59\pm0.01$)\\
    D-5 & 
    1.151100 (098) & 
    4.0 ($4.1\pm0.2$) & 
    11.70 ($11.70\pm0.02$)\\
    D-6 & 
    1.151200 (203) & 
    5.5 ($5.5\pm1.0$) & 
    11.60 ($11.60\pm0.06$)\\
    D-7 & 
    1.151270 (270) & 
    3.0 ($2.8\pm0.8$) & 
    11.50 ($11.47\pm0.08$)\\
    D-8 & 
    1.151350 (347) & 
    4.5 ($4.0\pm0.6$) & 
    11.80 ($11.77\pm0.05$)\\
    D-9 & 
    1.151420 (418) & 
    3.5 ($4.6\pm0.8$) & 
    11.60 ($11.69\pm0.06$)\\
    D-10 & 
    1.151500 (500) & 
    2.5 ($2.2\pm0.3$) & 
    12.00 ($12.00\pm0.02$)\\       
    D-11 & 
    1.151560 (562) & 
    5.0 ($5.3\pm0.7$) & 
    11.90 ($11.93\pm0.04$)\\ 
    D-12 & 
    1.151650 (651) & 
    4.0 ($3.9\pm0.1$) & 
    12.20 ($12.19\pm0.01$)\rule[-5pt]{0pt}{10pt}\\ 
    \hline
  \end{tabular}
\end{table}

We have synthesized four exemplary test cases to investigate the global
optimization performance of the CMA-ES when applied to the problem of
spectral decomposition. The test cases are based on the characteristics of
real QSO spectra and present a series of metal line ensembles increasing
in complexity. Test cases A, B, C, and D render superpositions of six,
eight, ten, and twelve components, respectively (Fig.~\ref{fg:cases}). All
test cases are synthesized simulating an instrumental resolution of
$R=60\,000$ and a curved background continuum.
Both Poissonian and Gaussian white noise are added producing a random
continuum noise level of about two percent. Since ES are not
destabilized by small scale oscillations of the objective function the
complexity of the decomposition problem does not depend on the noise level
of the data but is solely determined by the number and separation of
spectral lines. The parameterization of the artificial lines used to
synthesize the test cases is listed in Table~\ref{tb:cases}. 
Several lines are barely separated or occcur close to the limit of the
simulated spectral resolution.

The characteristics of the test cases are motivated by the analysis of QSO
spectra recorded with the UV-Visual Echelle Spectropgraph installed
at the Very Large Telescope. Further test cases exhibiting different
characteristics will emerge in the future in course of the analysis of QSO
spectra recorded with the Coud\'e Echelle Spectrograph operated at the
ESO~3.6~m telescope. We do not consider line ensembles consisting of less
than six components since these cases normally pose no challenge to the
CMA-ES algorithm. Primitive cases exhibiting just a single component are
completed in less than 50 generations using the canonical $(1,10)$-CMA-ES.

\subsection{Algorithm application}

We compare the global performance of the CMA-ES to that of several
classical optimization algorithms provided by the
\citetalias{PressTVF_2002} collection: Fletcher-Reeves-Polak-Ribiere
(FRPR, a conjugate gradient variant), Broyden-Fletcher-Gold\-fab-Shanno
(BFGS, a virtual metric variant), Levenberg-Mar\-quardt, Powell (a
direction set variant), and the downhill simplex. The two latter
algorithms do not require the calculation of partial derivatives and 
serve as direct standards of comparison. 

The positional search space is restricted to an interval being 40 percent
wider than the separation of the outer components (Fig.~\ref{fg:cases}),
while the broadening velocities and column densities are confined to 
$1.0\le b~(\mathrm{km\,s}^{-1})\le10.0$ and
$10.0\le\log N~(\mathrm{cm}^{-2})\le14.0$. Since the classical algorithms
are not prepared to handle simple parametric bounds per se, the
\citetalias{PressTVF_2002} routines are modified in such a way that any
step attempting to escape is suppressed. The partial derivatives required
by the FRPR, BFGS, and Levenberg-Marquardt algorithms are calculated
numerically using the symmetrized difference quotient approximation.

Regarding the CMA-ES, we  apply an $(100,200)$ algorithm with internal
strategy parameters preset to the canonical values except for
$\alpha_\mathrm{cov}=0$ and an increased covariance matrix adaption rate
$c_\mathrm{cov}$. The diagonal elements of $\mat{C}^{(0)}$ are initialized
to the squared half widths of the parameter intervals while the
mutation step size is initialized to $\sigma^{(0)}=0.5$.

For each algorithm and test case we monitor the history of 100 randomly 
initialized optimization runs. Providing the true number of absorption
components as prior information, the initial object parameters are
randomly drawn from a uniform distribution.
The random noise is generated only once for each test case and is the
same for all runs.
All optimization runs are stopped after 100\,000 evaluations of the
normalized RSS.

\subsection{Test results}

\begin{figure*}
  \centering
  \includegraphics{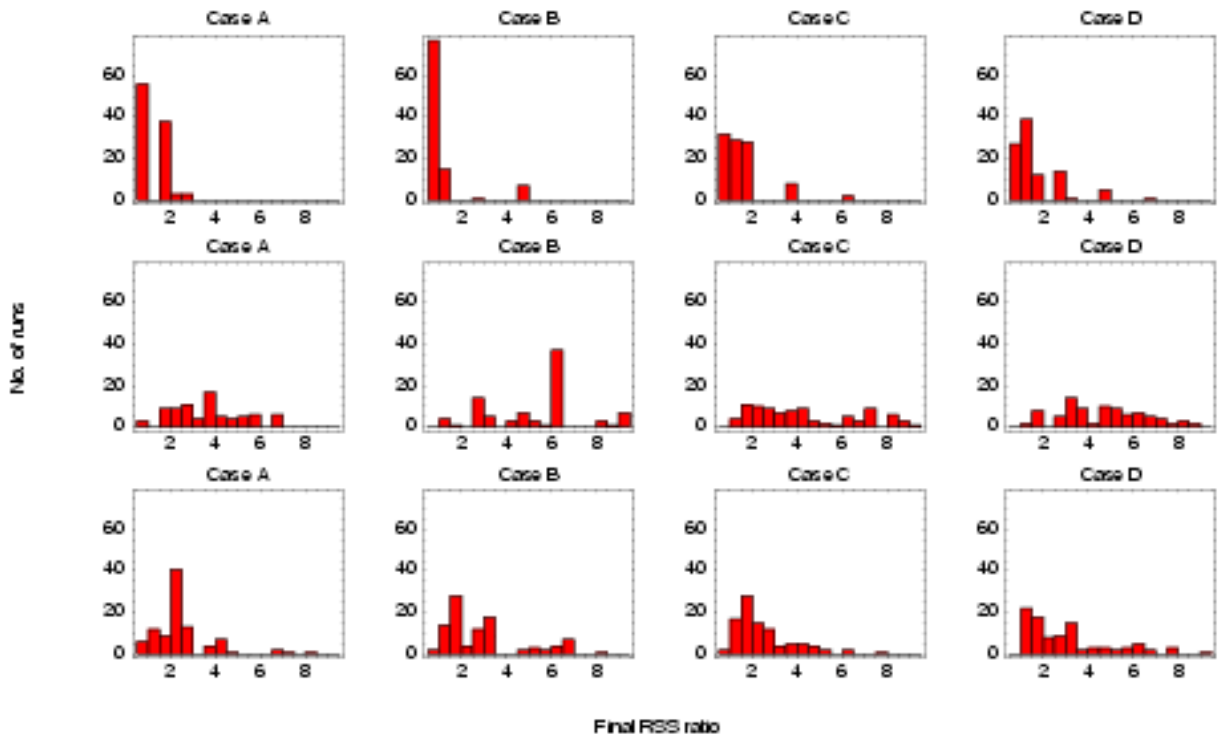}%
  \caption[]{Outcome of 100 randomly initialized optimization runs using
    the CMA-ES (first row of histograms), Levenberg-Marquardt (second
    row), and Powell algorithms.
    The number of runs converging at the global minimum is indicated by
    the leftmost bar.
    The FRPR, BFGS, and simplex algorithms perform worse and are omitted
    for convenience}
  \label{fg:hist}
\end{figure*}

\begin{figure*}
  \centering
  \includegraphics{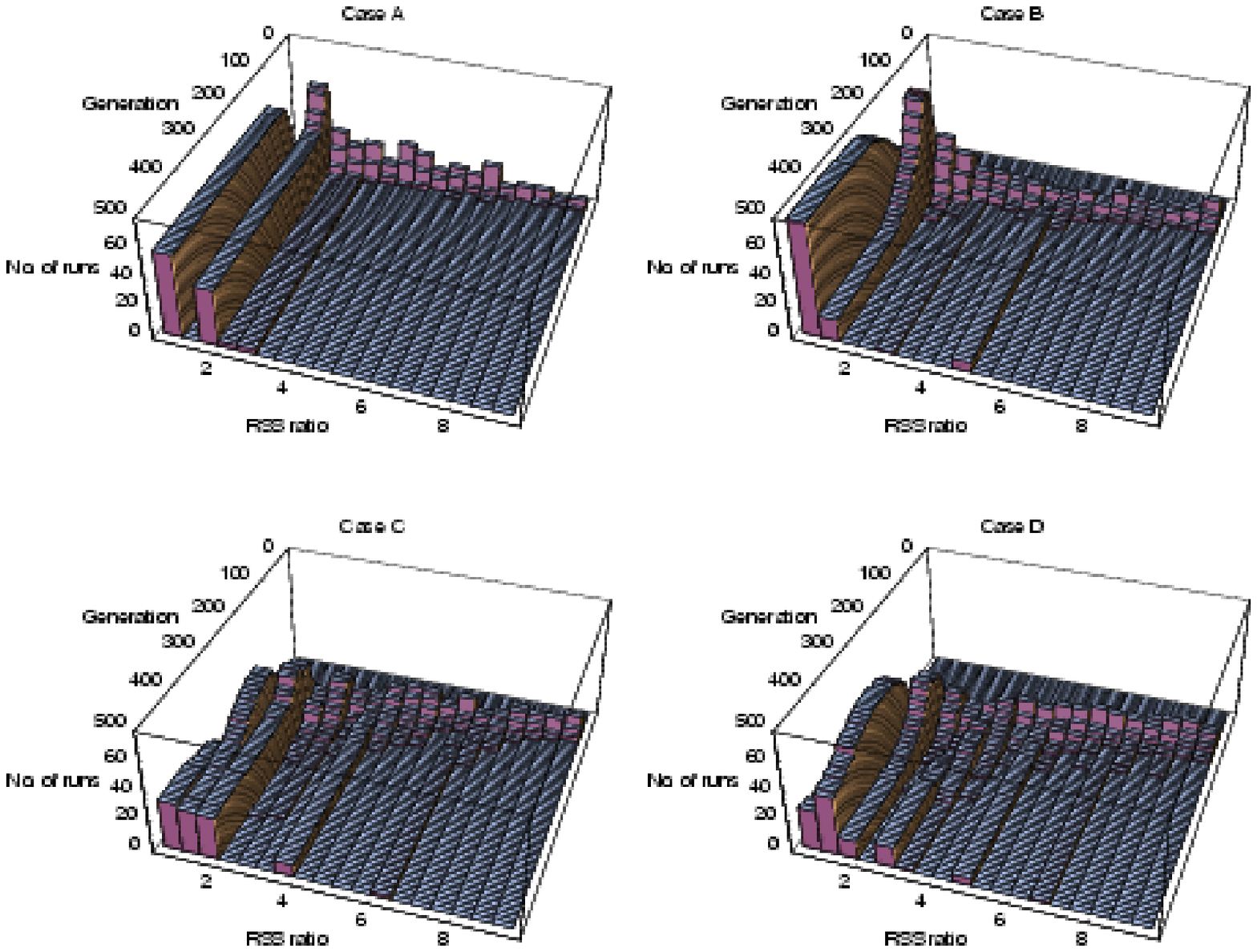}%
  \caption[]{History of 100 randomly initialized optimization runs using
    the CMA-ES.
    The number of runs converging at the global minimum is indicated by
    the leftmost bar.
    The number of RSS evaluations is calculated by multiplying the
    generation number with 200}
  \label{fg:hist3d}
\end{figure*}

The optimization runs are evaluated in terms of the ratio of the final
normalized RSS to the normalized RSS corresponding to the true
parameterization.
For all test cases the RSS ratio is marginally less than unity at the
global minimum. In general, an RSS ratio of unity just indicates the
statistical consistency of the optimized and the true absorption profiles,
but does not indicate the consistency of the optimized and the true
parameterization. In particular in the case of QSO spectra exhibiting
lower signal-to-noise ratios inconsistent optimized parameters will occur
regularly. But however, since in our test cases the true parameterization
is recovered by all runs converging at the global minimum 
(Table~\ref{tb:cases}) the RSS ratio is appropriate to assess the
global optimization performance.
The outcome of optimization runs is summarized in Fig.~\ref{fg:hist}
and Table~\ref{tb:stat} while Fig.~\ref{fg:hist3d} illustrates the
performance of the CMA-ES during different stages of the optimization. In
the following paragraphs the test results are described in detail.

\begin{table*}
  \centering
  \caption[]{Global optimization performance summarized in terms of the
    median of the final normalized RSS ratio, the number of runs
    converging at the global minimum, and the median number of RSS
    evaluations needed for that}
  \label{tb:stat}
  \begin{tabular}{@{}l
    l@{\hspace{0.5\columnsep}}l@{\hspace{0.5\columnsep}}l
    l@{\hspace{0.5\columnsep}}l@{\hspace{0.5\columnsep}}l
    l@{\hspace{0.5\columnsep}}l@{\hspace{0.5\columnsep}}l
    l@{\hspace{0.5\columnsep}}l@{\hspace{0.5\columnsep}}l@{}}
    \hline
    \hline
    Algorithm &
    \multicolumn{3}{l}{Case A}\rule[-5pt]{0pt}{15pt} &
    \multicolumn{3}{l}{Case B} &
    \multicolumn{3}{l}{Case C} &
    \multicolumn{3}{l}{Case D}\\
    \hline
    CMA-ES\rule{0pt}{10pt} &
    0.97 & 56 & 18\,000 &
    0.99 & 77 & 40\,000 &
    1.04 & 32 & 62\,000 &
    1.09 & 27 & 82\,000 \\
    FRPR &
    4.86 & 2 & 36\,000 &
    9.45 & 3 & 70\,000 &
    6.65 & -- & -- &
    11.7 & -- & -- \\
    BFGS &
    26.8 & -- & -- &
    32.9 & -- & -- &
    34.7 & -- & -- &
    37.5 & -- & -- \\
    Levenberg-Marquardt &
    3.97 & 3 & 1000 &
    6.11 & -- & -- &
    4.24 & -- & -- &
    5.04 & -- & -- \\
    Powell &
    2.12 & 6 & 14\,000 &
    2.67 & 2 & 20\,000 &
    2.05 & 2 & 26\,000 &
    2.69 & -- & -- \\
    Simplex\rule[-5pt]{0pt}{10pt} &
    12.2 & 2 & 68\,000 &
    25.6 & -- & -- &
    6.96 & -- & -- &
    35.8 & -- & -- \\
    \hline
  \end{tabular}
\end{table*}

\subsubsection{Case A}

Case A consists of four isolated and two blended components. The lowest
local minimum occurs when the weaker line of the blend is missed, but an
isolated line is hit twice instead. Higher local minima occur whenever the
blend is modelled correctly but any isolated component is missed. The
CMA-ES hits the global minimum in 56 and the lowest local minimum in 38
runs. Another six runs converge at higher local minima while missing an
isolated component. The most stable classical technique is the Powell
algorithm, arriving at the global minimum in six, at the lowest local
minimum in nine, and between these both in twelve runs. The most
successful gradient technique is the Levenberg-Marquardt algorithm,
reaching the global minimum in three runs. The latter algorithm also is
the fastest, needing about 1000 evaluations of the normalized RSS to
locate the global minimum, whereas the Powell and CMA-ES algorithms
require about 14\,000 and 18\,000 evaluations, respectively. The FRPR and
simplex algorithms hit the global minimum in two runs each, needing about
36\,000 and 68\,000 RSS evaluations, respectively.
Except for the Powell algorithm the majority of deterministic runs misses
two or more components. The BFGS algorithm never hits any component.

\subsubsection{Case B}

Case B is similar to case A but exhibits two additional components and a
less narrow blend. The CMA-ES hits the global minimum in 77 and the lowest
local minimum in 15 runs. Another eight runs converge at higher local
minima. No deterministic algorithm locates the global minimum in more than
three runs. The Powell algorithm is still relatively stable, missing one
or two components in the majority of runs.

\subsubsection{Case C}

Case C again exhibits two additional components resulting in an ensemble
of largely blended lines. The lowest local minimum occurs when the seventh
component is missed, but another component is hit twice instead. The
global minimum of the objective function almost has degenerated, being
different from the lowest local minimum merely by seven percent. The
CMA-ES hits the global minimum in 32 and the lowest local minimum in 22
runs. The majority of the remaining runs converges at higher local minima
while missing a different one than the seventh component. The Powell
algorithm hits the global as well as the lowest local minimum in 2 runs
each and misses one or two components in the majority of runs. No further
classical algorithm hits the global minimum.

\subsubsection{Case D}

Case D is very similar to case C but exhibits two additional components.
The CMA-ES hits the global minimum in 27 runs. The majority of runs
arrives at lower local minima without becoming stationary
(Fig.~\ref{fg:hist3d}). No classical algorithm hits the global minimum.

\subsubsection{Efficiency}
The number of RSS evaluations required by the CMA-ES to converge at the
global minimum increases linearly with the number of lines superimposed
in the test case (Table~\ref{tb:stat}). The linear correlation
coefficient is about $a=10\,000$ RSS evaluations per line. In case of the
Powell algorithm the linear correlation coefficient is about $a=3000$.

\section{Summary and conclusions} 

All tested classical algorithms require an adequate initial guess to
locate the global optimum of the objective function when applied to the
problem of spectral decomposition. In contrast, the CMA-ES is demonstrably
capable to calculate the optimal decomposition without demanding any
particular initialization, and therefore is just asking to be used in
automatized spectroscopic analysis software.

The CMA-ES does not guarantee to find the optimal solution: characteristic
spectral features are modelled correctly, but features exhibiting just a
small attractor volume such as narrow components or tight blends are
frequently not distinguished fittingly. Larger populations are expected to
improve the chance to hit the global optimum but require a greater number
of objective function evaluations. Peak finding algorithms could detect
both the proper number (a problem that we have completely left aside) and
position of spectral components and provide an improved initialization. 
But irrespective of whether an automatized analysis software will be
realized, the CMA-ES is an elegant and highly competitive general
purpose algorithm which is easy to implement. In our opinion, its
integration into the standard astronomical data analysis packages will be
thoroughly worthwhile
and its widespread use will contribute to the further comprehension and
improvement of suchlike algorithms. 

\begin{acknowledgements}
  We kindly thank Nikolaus Hansen for stimulating and valuable
  correspondence and for reading the manuscript.
  This research has been supported by the Verbundforschung of the BMBF/DLR
  under Grant No.~50~OR~9911~1.
\end{acknowledgements}

\bibliographystyle{aa}
\bibliography{abbrev,astron,num,rq}

\end{document}